\documentclass[a4paper,12pt]{article}

\usepackage{amsfonts,amssymb,amsmath,amsthm}

\numberwithin{equation}{section}

\theoremstyle{plain}
\newtheorem{theorem}{Theorem}[section]

\newtheorem{lemma}[theorem]{Lemma}

\newtheorem{remark}[theorem]{Remark}

\DeclareMathOperator{\curl}{curl}

\DeclareMathOperator{\dist}{dist}

\newcommand{\ball}[2]{{\cal B}_{#2}\left(#1\right)}

\newcommand{\bydef}{\ensuremath{:=}}

\newcommand{\D}{\ensuremath{\, d}}

\newcommand{\imunit}{\mathrm{i}}

\newcommand{\domain}{\mathcal{D}}

\newcommand{\essspectrum}{\sigma_{ess}}

\newcommand{\ie}{i.e. }

\newcommand{\Ltwo}{L^2}

\newcommand{\real}{\mathbb R}
\newcommand{\sobspace}{H^1}
\newcommand{\spectrum}{\sigma}

\newcommand{\form}{\mathfrak{q}}

\def\Om{\Omega}
\def\om{\omega}

\def\g{\gamma}
\def\G{\Gamma}
\def\l{\lambda}
\def\p{\partial}

\def\a{\alpha}
\def\b{\beta}
\def\t{\widetilde}

\begin{document}

\noindent{\large\textbf{Spectrum of the Magnetic Schr\"odinger
Operator in a Waveguide with Combined Boundary Conditions}}

\bigskip

\noindent{\large Denis~Borisov, Tomas~Ekholm and Hynek~Kova\v
r\'{\i}k}

\begin{abstract}
We consider the magnetic Schr\"odinger operator in a
two-dimensional strip. On the boundary of the strip the
Dirichlet boundary condition is imposed except for a fixed
segment (window), where it switches to magnetic
Neumann\footnote{For the definition of magnetic Neumann boundary
conditions see Section 2, Eq. (\ref{mgnm})}. We deal with a
smooth compactly supported field as well as with the
Aharonov-Bohm field. We give an estimate on the maximal length
of the window, for which the discrete spectrum of the considered
operator will be empty. In the case of a compactly supported
field we also give a sufficient condition for the presence of
eigenvalues below the essential spectrum.
\end{abstract}

\section{Introduction}
The existence of bound states of the Laplace operator in the
strip with Dirichlet boundary conditions and Neumann ``window''
was proven in \cite{BGRS} and independently also in \cite{ESTV}.
The so called Neumann window is represented by the segment of
the length $2l$ of the boundary, on which the Dirichlet
conditions are changed to Neumann. A discrete spectrum of the
Laplace operator with Neumann window appears for any nonzero
length of the Neumann segment. In particular, for small values
of $l$ the eigenvalue emerges from the continuous spectrum
proportionally to $l^4$. The asymptotical estimate for small $l$
were established in  \cite{EV}, while the rigorous results on
asymptotical expansions were obtained in \cite{G}.

On the other hand, the results on the discrete spectrum of a
magnetic Schr\"odinger operator in waveguide-type domains are
scarce. A planar quantum waveguide with constant magnetic field
and a potential well is studied in \cite{DEM}, where it was
proved that if the potential well is purely attractive, then at
least one bound state will appear for any value of the magnetic
field.
\par In this work we consider the system, where the discrete
spectrum in the absence of magnetic field appears due to the
perturbation of the boundary of the domain rather than due to
the additional potential well. We also assume that the magnetic
field is localised in the sense to be specified below. This
assumption rules out the case of a constant field. As it has
been recently shown in \cite{EK} the presence of a suitable
magnetic field can prevent the existence of bound states in the
Dirichlet strip with a sufficiently small ``bump''. Changing the
boundary conditions to Neumann is however a stronger
perturbation in the sense that the existence of a bound state in
a waveguide with the bump added to a certain segment of the
boundary implies the existence of a bound state in a waveguide
with Neumann conditions on the same segment, see \cite[Cor.
1.3]{BGRS}. Therefore we cannot mimick the arguments of
\cite{EK} in the case of  the waveguide with Neumann window and
a different approach is needed. \par The main technical tool
used in \cite{EK} is a modified version of the Hardy inequality
for the magnetic Dirichlet quadratic form in the two-dimensional
strip. In the present paper we establish a similar inequality in
order to prove the absence of a discrete spectrum of the
magnetic Schr\"odinger operator in the straight strip with
Neumann window. More exactly speaking, we give sufficient
conditions on the magnetic field and the length of the window,
under which the discrete spectrum is empty. The above mentioned
version of Hardy inequality enables us to reduce the problem to
the study of a one-dimensional Laplacian with a purely
attractive potential well of a width $2l$ and a small but fixed
positive potential, see Section 4.2 for the details. We then
show that for $l$ small enough such a system has no bound state.
The main profit of our method is that it gives us an explicit
estimate on the critical length of the window, depending on the
magnetic field, which guarantees the absence of discrete
spectrum.

It is of course natural to ask  whether a sufficiently large
Neumann window will lead to the existence of eigenvalues also in
the presence of the magnetic field. In the case of a smooth and
compactly supported field we give an answer to this question
using a minimax-like argument.

The article is organised as follows. In Section 2 we define the
mathematical objects that we work with and describe the problem.
We also give the statements of the main results separately for
the case of a compactly supported bounded magnetic field and for
the Aharonov-Bohm field. In Section 3 we show that the essential
spectrum of the Dirichlet Laplacian is not affected by the
magnetic field, neither by the presence of Neumann window.
Sufficient conditions for the absence of the discrete spectrum
are proved in Section 4. Finally, the question of presence of
eigenvalues is discussed in Section 5.

\section{Statement of the problem and the main results}

Let $x=(x_1,x_2)$ be Cartesian coordinates, $\Omega$ be the
strip $\{x: 0<x_2<\pi\}$, and $\gamma$ be the interval $\{x:
|x_1|<l, x_2=0\}$. The rest of the boundary will be indicated by
$\Gamma$, \ie $\Gamma = \p\Omega\setminus\overline{\gamma}$. We
denote by $B=B(x)$ a real-valued magnetic field and assume that
$A$ is a magnetic vector potential associated with $B$, \ie
$A=A(x)=(a_1(x),a_2(x))$ and $B=\curl A= \p_{x_1}a_2 - \p_{x_2}
a_1$. In what follows we will consider two main cases of
magnetic fields $B$. The first case is a smooth compactly
supported field. Hereinafter by this we denote the field $B$
belonging to $C^1(\overline{\Om})$ and vanishing in the
neighbourhood of infinity. The second one is the Aharonov-Bohm
field originated by the potential with components
\begin{equation}\label{ahbm}
a_1(x)=-\frac{\Phi \cdot (x_2-p_2)}{(x_1-p_1)^2+(x_2-p_2)^2},
\quad a_2(x)=\frac{\Phi \cdot
(x_1-p_1)}{(x_1-p_1)^2+(x_2-p_2)^2},
\end{equation}
where $\Phi$ is a constant and $2\pi\Phi$ is the flux through
the point $p=(p_1,p_2)$ which is assumed to be inside the strip
$\Om$. We denote by $M_0$ the operator
\begin{equation*}
\left(-\imunit\p_{x_1} + a_1\right)^2 + \left(-\imunit\p_{x_2} +
a_2 \right)^2
\end{equation*}
on the domain $\domain (M_0)$ consisting of all functions $u \in
C^\infty(\overline{\Omega})$ vanishing in a neighborhood of
$\Gamma$ and in a neighborhood of infinity and satisfying the
boundary condition
\begin{equation}\label{mgnm}
(-\imunit\p_{x_2}+a_2){u}(x)=0\quad\text{on $\gamma$}.
\end{equation}
We will call it magnetic Neumann boundary condition. In the case
of Aharonov-Bohm field, the functions $u\in\domain (M_0)$ are
assumed to vanish in a neighbourhood of the point $p$. Clearly,
the operator $M_0$ is non-negative and symmetric in $\Ltwo(\Om)$
and therefore it can be extended to a self-adjoint non-negative
operator by the method of Friedrich. In what follows we will
denote this extension by $M$. The main object of our interest is
the spectrum of the operator $M$.

In order to formulate the main results we need to introduce some
auxiliary notations. By $\Om (\a,\b)$ we will indicate the
subset of $\Om$ given by $\{x \in \Om : \a<x_1< \b \}$ and
$\Omega_\pm$ will be the subsets $\left\{x \in \Om : x_1 > l
\right\}$, $\left\{ x \in \Om : x_1 < -l\right\}$, respectively.
The symbol $\ball{q}{r}$ denotes a ball of radius $r$ centered
at a point $q$ in $\real^2$. The flux of the field through the
ball $\ball{q}{r}$ is given by
\begin{equation*}
\Phi_q(r) = \frac{1}{2\pi}\int_{\ball{q}{r}}B(x)\D x.
\end{equation*}



Below we give the summary of the  main results of the article.

\begin{theorem}\label{th0}
The essential spectrum of the operator $M$ coincides with
$[1,+\infty)$.
\end{theorem}

\begin{theorem}\label{th1}
Assume that the field $B$ is smooth and compactly supported and
\begin{enumerate}
\def\theenumi{(\arabic{enumi})}
\item\label{cnA0}
There exist two balls $\ball{p_-}{R_-}\subset\Om_-$,
$\ball{p_+}{R_+}\subset\Om_+$ so that at least one of the fluxes
$\Phi_{p_\pm}(r)$ is not identically zero for $r\in[0,R_\pm]$;

\item\label{cnA1}
The inequality
\begin{equation}
\begin{aligned}
l\le \frac{1}{12}\left(\kappa_-+\kappa_+\right)
\end{aligned}\label{1.1}
\end{equation}
holds true, where
\begin{equation}\label{1.10}
\kappa_\pm\bydef\min\left\{\pi c_\pm,\frac{\pi}{4\ln 2
+\pi|p_1^\pm|}\right\},
\end{equation}
$c_\pm$ are defined in Lemma~\ref{hardylemma}.
\end{enumerate}
Then the operator $M$ has empty discrete spectrum.
\end{theorem}

\begin{theorem}\label{th3}
Assume that the field $B$ is the Aharonov-Bohm one with the
potential given by (\ref{ahbm}) and
\begin{enumerate}
\def\theenumi{(\arabic{enumi})}
\item\label{cnA2}

The point $p$ is $(p_1,p_2)$, where $p_1<-l$;

\item\label{cnA3}
The inequality
\begin{equation}
l<\frac{\kappa}{6}\label{1.3}
\end{equation}
holds true, where
\begin{equation}\label{1.11}
\kappa\bydef\min\left\{\pi c,\frac{\pi}{4\ln 2+\pi
|p_1|}\right\},
\end{equation}
$c$ is defined in Lemma~\ref{hardylemma2}.
\end{enumerate}
Then the operator $M$ has empty discrete spectrum.
\end{theorem}

The next theorem provides a condition, that guarantees the existence
of discrete eigenvalues in the case of a smooth and compactly
supported field.


\begin{theorem}\label{th2}
Let the field $B$ be smooth and compactly supported, $\l=\l(l)$
be the lowest eigenvalue of the Laplacian
$-\Delta_{\mathcal{N},\mathcal{D}}$ in the strip $\Om$ subject
to the Dirichlet condition on $\G$ and Neumann condition on
$\g$. Assume that the inequality
\begin{equation}
\l(l)+\inf\limits_{A}\max\limits_{\overline{\Om}}|A(x)|^2<1\label{1.2}
\end{equation}
holds, where infimum is taken over all potentials associated
with the field $B$. Then the operator $M$ has non-empty discrete
spectrum.
\end{theorem}

\begin{remark}\label{rm1.1}
It will be shown in the proof of Theorem~\ref{th2} that under
the hypothesis of this theorem the potential $A$ can be chosen
such that $|A|$ is bounded and of compact support. This will
imply that  the quantity
$\inf\limits_{A}\max\limits_{\overline{\Om}}|A(x)|^2$ in
(\ref{1.2}) is finite.
\end{remark}

Throughout the article we will often make use of some notations
and it is convenient to introduce them now. The spectrum of an
operator $T$ will be indicated by $\spectrum(T)$ while the
essential spectrum will be denoted by $\essspectrum(T)$. We will
employ the symbol $\form_T=\form_T[\cdot,\cdot]$ for the
sesquilinear form associated with a self-adjoint operator $T$
and $\domain(\form_T)$ will be the domain of the quadratic form
produced by the sesquilinear form $\form_T$. The Hilbert space
we will work in is $\Ltwo(\Om)$; we preserve the notation
$(\cdot,\cdot)$ and $\|\cdot\|$ for the inner product and norm
in this space. In all other cases the notations of the inner
product and norm in a Hilbert space $H$ will be equipped by a
subscript $H$.


\section{Proof of Theorem~\ref{th0}}

To prove the theorem we will need some auxiliary notations and
statements. Let $H$ be a Hilbert space and $S$ be a positive
definite operator in $H$ whose domain is dense in $H$. By $S_1$
we indicate the Friedrich's extension of the operator $S$ and by
$S_2$ another self-adjoint positive definite extension of $S$.
By definition, $\domain(\form_{S_2})$ is a Hilbert space endowed
with the inner product and the norm originated by the quadratic
form $\form_{S_2}$. Since $S_1$ is the Friedrich's extension of
$S$ it follows that $\domain(\form_{S_1})$ is a subspace of
$\domain(\form_{S_2})$. Let $\mathcal{Q}$ be the orthogonal
complement $\domain(\form_{S_1})^\bot$ in $\domain(\form_{S_2})$
in the inner product $\form_{S_2}[\cdot,\cdot]$.

The proof of the theorem is based on the following lemma proven
in \cite[Lemma 3.1]{B}.

\begin{lemma}\label{lm6.1} 
If each bounded subset of $\mathcal{Q}$ (in the norm
$\|\cdot\|_{\domain(\form_{S_2})}$) is compact in $H$,  then the
operator $T\bydef S^{-1}_2-S_1^{-1}$ is compact in $H$.

%

\end{lemma}

In our case $\Ltwo(\Om)$ plays the role of $H$ and
$S\bydef(-\imunit\nabla+A)^2+1$ with $\domain(S)\bydef
C^\infty_0(\Om)$. The Friedrich extension $S_1$ of $S$ is in
fact the extension of $(-\imunit\nabla+A)^2+1$ subject to
Dirichlet boundary condition. We know from \cite{EK} that
$\essspectrum(S_1)=[2,+\infty)$. We set $S_2\bydef M+1$; we
naturally can treat $M+1$ as an extension of $S$. If we prove
that $T\bydef S_2^{-1}-S_1^{-1}$ is compact, then the essential
spectra of the operators $S_1$ and $S_2$ will coincide by the
Weyl theorem (see for instance \cite{BS1}). We will prove the
compactness of $T$ by Lemma~\ref{lm6.1}. First we will establish
an auxiliary lemma. By $\om$ we indicate some bounded subdomain
of $\Om$ with
infinitely differentiable boundary such that 
$\dist(\gamma,\Om\setminus\overline{\om})>0$. In the case of
Aharonov-Bohm field we also
assume that the 
point $p$ 
does not belong to $\om$.

\begin{lemma}\label{lm6.2}
For each function $u\in\mathcal{Q}$  the inequality
\begin{equation*}
\|u\|\le c\|u\|_{\Ltwo(\om)},
\end{equation*}
holds true, where the constant $c$ is independent on $u$.
\end{lemma}
\begin{proof}
In the proof of the lemma we follow the
ideas of the proof of Lemma 3.3 in \cite{B}. 
The domains $\domain(\form_{S_1})$ and $\domain(\form_{S_2})$
are completions of $C_0^\infty(\Om)$ and $\domain(M_0)$,
respectively, in norm
\begin{equation*}
\|(-\imunit\nabla+A)\cdot\|^2+\|\cdot\|^2.
\end{equation*}
In the case of compactly supported field we can choose the vector potential $A$
being from $C^1(\overline{\Om})$ which will make this potential
bounded on $\overline{\om}$. In the case of Aharonov-Bohm field
the potential is in $C^1(\overline{\om})$ as well since the
point $p$ does not belong to $\om$ by assumption. Therefore,
each element $v$ of $\domain(S_2)$ belongs to $\sobspace(\om)$
due to the inequality:
\begin{equation}
\begin{aligned}
\| v\|_{\sobspace(\om)}^2&=
\|(-\imunit\nabla+A)v-Av\|_{\Ltwo(\om)}^2+\|v\|^2_{\Ltwo(\om)}
\\
&\le 2\left(\|(-\imunit\nabla+A)v\|^2_{\Ltwo(\om)}+\|A
v\|^2_{\Ltwo(\om)}\right)+\|v\|^2_{\Ltwo(\om)}
\\
&\le c\left(\|(-\imunit\nabla+A)v\|^2_{\Ltwo(\om)}+
\|v\|^2_{\Ltwo(\om)}\right)=c(S_2v,v),
\end{aligned}\label{6.1}
\end{equation}
where the constant $c$ is independent on $v$.

We denote by $\chi=\chi(x)$ an infinitely differentiable
function taking values from $[0,1]$ and being equal to
one in some neighbourhood of $\gamma$, which is a subdomain of
$\om$, and vanishing outside $\om$. Since $S_2\ge 1$ it follows
that
\begin{equation}\label{6.3}
\|S_2^{-1}u\|\le \|u\|.
\end{equation}
Let $u\in\mathcal{Q}$. Clearly, $(1-\chi)S_2^{-1}u\in
\domain(\form_{S_1})\cap\domain(S_2)$, thus
\begin{equation*}
\left(S_2(1-\chi)S_2^{-1}u,u\right)=\left((1-\chi)
S_2^{-1}u,u\right)_{\domain(\form_{S_2})}=0.
\end{equation*}
Using this equality we deduce
\begin{equation}\label{6.2}
\|u\|^2=(u,u)-\left(S_2(1-\chi) S_2^{-1}u,u\right)=(S_2\chi
S_2^{-1}u,u).
\end{equation}
Since
\begin{equation*}
S_2\chi S_2^{-1}u=\chi u-2\left(\nabla
(S_2^{-1}u),\nabla\chi\right)_{\real^2}-(S_2^{-1}u)\Delta\chi-
2\,\imunit \left(A,\nabla\chi\right)_{\real^2} S_2^{-1}u
\end{equation*}
due to (\ref{6.1})--(\ref{6.2}) we have
\begin{align*}
\|u\|^2&\le\int_\Om\chi|u|^2\D x
+c\|u\|_{\Ltwo(\om)}\|S_2^{-1}u\|_{\sobspace(\om)} \\ &\le
c\|u\|_{\Ltwo(\om)}\left(\|u\|+\sqrt{(S_2^{-1}u,u)}\right)\le
c\|u\|_{\Ltwo(\om)}\|u\|,
\end{align*}
where $c$ is independent on $u$. This proves
the lemma.
\end{proof}


Let us finish the proof of the Theorem. Given a subset $K$ of
$\mathcal{Q}$ bounded in the norm
$\|\cdot\|_{\domain(\form_{S_1})}$, we conclude that it is also
bounded in $\sobspace(\om)$ due to (\ref{6.1}). By the well
known theorem on compact embedding of $\sobspace(\om)$ in
$\Ltwo(\om)$ for each bounded domain with smooth boundary (see,
for instance, \cite{L}) we have that the set $K$ is compact in
$\Ltwo(\om)$. Applying now Lemma~\ref{lm6.2}, we conclude that
$K$ is compact in $\Ltwo(\Om)$. Hence, the assumption of
Lemma~\ref{lm6.1} is satisfied and operator $T$ introduced above
is compact. The proof of Theorem~\ref{th0} is complete.

\section{Absence of the discrete spectrum}

This section is devoted to the proof of Theorems~\ref{th1}
and~\ref{th3}. By Theorem~\ref{th0} we know that the essential
spectrum of the operator $M$ is $[1,+\infty)$. Thus, the
equivalent formulation of the absence of the discrete spectrum
is the following inequality
\begin{equation}
\inf\spectrum(M-1) = \inf_{
\genfrac{}{}{0 pt}{1}{\|u\| = 1}{u \in \domain(\form_{M})}
}\left(\|(-\imunit\nabla+A)u\|^2-\|u\|^2\right)\ge 0.
\end{equation}
It will be enough to check the infimum for a
$\|\cdot\|_{\domain(\form_M)}$-dense subset of $\domain(M)$. Hence
\begin{equation}
\inf\spectrum(M-1) = \inf_{
\genfrac{}{}{0 pt}{1}{\|u\| = 1}{u \in \domain (M_0)}
}\left(\|(-\imunit\nabla+A)u\|^2-\|u\|^2\right)\ge 0 \label{4.1}
\end{equation}
 In order to prove this we will need some auxiliary statements which will be
 established in the next two subsections.

\subsection{A Hardy inequality}

Here we state a Hardy inequality for the quadratic form of the
operator $M$, which will be one of the crucial tools in the
proofs of Theorems~\ref{th1} and~\ref{th3}. Let $p=(p_1,p_2)\in\Om$ be some point
and the number $R$ be such that $\ball
{p}{R}\subset\overline{\Om}$. Given a smooth compactly supported
field $B$, we define the function $\mu(r)\bydef
\dist(\Phi_p(r),\mathbb{Z})$, where we recall that $\Phi_p(r)$
is the flux of the field $B$ through the ball $\ball {p}{r}$.
We introduce the function
\begin{equation}\label{2.5}
c(p,R)  =\left\{
\begin{aligned}
&\frac{1}{16+c_1(R) c_2(p,R)},&&\text{if $\Phi_p(r)\not\equiv0$
as $r\in[0,R]$},
\\
&0,&&\text{if $\Phi_p(r)\equiv0$ as $r\in[0,R]$},
\end{aligned}\right.
\end{equation}
where
\begin{equation}\label{3.13}
\begin{aligned}
&c_1(R) =\frac{64+4R^2}{R^4},
\\
&c_2(p,R)= \frac{2R^2 c_3(p_2) c_4(R) + 4c_4(R) + 4R^2}{
c_3(p_2) \, \cos^2(|p_2 - \frac{\pi}{2}| + R)},
\\
&c_3(p_2)=\pi^2\min\{p_2^{-2},(\pi-p_2)^{-2}\}-1,
\\
& c_4(p,R) =
\max_{[0,R]}\left|\left(\frac{\mu(r)}{r}\right)'\right|,
\\
&c_5(R)= \max \left\{ 2 \mu_0^2 + 4 c_5^2 c_6\mu_0^4, c_6
\right\},
\\
& c_6(R)=4 \max \left\{ \frac{r_0^2}{\nu_0^{2}},\frac{ 2R^3 - 3
R^2 r_0 + r_0^3}{6r_0} \right\}
\end{aligned}
\end{equation}
and $\mu_0$ and $r_0$ are defined by
\begin{equation*}
\mu_0\bydef\frac{1}{\max\limits_{[0,R]}
r^{-1}\mu(r)}v=\frac{r_0}{\mu(r_0)},
\end{equation*}
$\nu_0$ is a smallest positive root of the Bessel function
$J_0$.

%

%
It was shown in \cite{EK} that the function $c(p,R)$ is well
defined. Finally, let us define
\begin{equation}\label{g}
g(x_1)=\left\{\begin{aligned} &1,&&\text{if }|x_1|>l,
\\
&\frac{1}{4},&&\text{if }|x_1|\le l.
\end{aligned} \right.
\end{equation}




\begin{lemma} \label{hardylemma}
Assume that the field $B$ is smooth and compactly supported and
the condition \ref{cnA0} of Theorem~\ref{th1} is satisfied for
the points $p_-=(p_1^-, p_2^-)$ and $p_+=(p_1^+, p_2^+)$, then
\begin{equation}\label{3.1}
\int_\Omega \rho(x_1) |u|^2 \D x \leq \int_\Omega  \left( |(- i
\nabla + A)
  u|^2 -  g(x_1)|u|^2 \right) \D x ,
\end{equation}
holds for all $u \in \domain (M_0)$, where
\begin{equation}\label{3.2}
\rho (x_1) =\left\{
\begin{aligned}
&\frac{c_-}{1+(x_1-p_1^-)^2},\quad&&\text{if }-\infty<x_1<p_1^-,
\\
&\hphantom{1+(x_1}0,\quad&&\text{if } p_1^-<x_1<p_1^+,
\\
&\frac{c_+}{1+(x_1-p_1^+)^2},\quad&&\text{if }
p_1^+<x_1<+\infty,
\end{aligned}
\right.
\end{equation}
and the constants $c_\pm = c(p_\pm,R_\pm)$ are given by (\ref{2.5}).
\end{lemma}
\begin{proof} We start the proof from the estimate
\begin{equation} \label{hardy1}
c_- \int_{\Omega(-\infty,p_1^-)} \frac{|u|^2}{1 + (x_1 -
p_1^-)^2} \D x \leq \int_{\Omega(-\infty,p_1^-)} \left(
|(-\imunit \nabla + A)u|^2 - |u|^2 \right) \D x,
\end{equation}
which is valid for all $u\in\domain(M_0)$. The proof of this
estimate follows from the calculations of \cite[Sec. 6]{EK},
where the similar inequality
\begin{equation}\label{3.11}
c\int_{\Omega} \frac{|u|^2}{1 + (x_1 - p_1^-)^2} \D x \leq
\int_{\Omega} \left( |(-\imunit \nabla + A)u|^2 - |u|^2 \right)
\D x,
\end{equation}
is proved for all $u\in \sobspace_0(\Om)$ with some constant
$c$. The approach employed in \cite[Sec. 3]{EK} can be applied
to prove the inequality (\ref{hardy1}). We will not reproduce
all the details of this proof and just note that the only
modification needed is to replace the function $\varphi$ defined
in \cite[Eq. (3.28)]{EK} by
\begin{equation}\label{3.12}
\varphi(x):= \left\{
\begin{aligned}
&1 && \text{ if $x_1
< p_1^- - \frac {R}{\sqrt 2}$},
\\
&\frac{\sqrt{2}(p_1^- - x_1)}{R} && \text{ if } p_1^- - \frac
R{\sqrt 2} < x_1 < p_1^-,
 \\
& 0 && \text{ elsewhere,}
\end{aligned} \right.
\end{equation}
In the same way the inequality
\begin{equation} \label{hardy2}
c_+ \int_{\Omega(p_1^+,+\infty)} \frac{|u|^2}{1 + (x_1 -
p_1^+)^2} \D x \leq \int_{\Omega(p_1^+,+\infty)} \left(
|(-\imunit \nabla + A)u|^2 - |u|^2 \right) \D x,
\end{equation}
holds for all $u \in \domain(M_0)$, where $c_+ = c(p_+,R_+)$. We
will make use of the diamagnetic inequality (see \cite{HS})
\begin{equation}\label{diam}
|\nabla |u|(x)| \leq |(-\imunit\nabla + A)u(x)|
\end{equation}
which holds pointwise almost everywhere in $\Omega$ for each $u
\in \domain (M_0)$. In addition the trivial inequality
\begin{equation}\label{trin}
\int_0^\pi |\p_{x_2}u|^2 \D x_2 \geq \int_0^\pi g|u|^2 \D x_2
\end{equation}
holds for each fixed $x_1$ and all $u\in \domain (M_0)$. The
diamagnetic inequality (\ref{diam}) and the last estimate lead
us to the inequality
\begin{align*}
\int_{\Omega(\alpha,\beta)} \left( |(-\imunit \nabla +
A)u|^2\right) \D x \ge\int_{\Omega(\alpha,\beta)} |\nabla|u||^2
\D x \ge\int_{\Omega(\alpha,\beta)} g|u|^2 \D x,
\end{align*}
which is valid for all $\alpha < \beta$.  Combining now this
inequality with (\ref{hardy1}), (\ref{hardy2}) we arrive at the
statement of the lemma.
\end{proof}

In the case of the Aharonov-Bohm field the similar statement is
true.

\begin{lemma} \label{hardylemma2}
Assume that the field is generated by Aharonov-Bohm potential
given by (\ref{ahbm}) and that the condition \ref{cnA2} of the
theorem~\ref{th3} is satisfied for the point $p=(p_1,p_2)$. Then
\begin{equation}\label{3.3}
\int_\Omega \rho(x_1) |u|^2 \D x \leq \int_\Omega  \left( |(- i
\nabla + A)
  u|^2 -  g(x_1)|u|^2 \right) \D x ,
\end{equation}
holds for all $u \in \domain (M_0)$, where
\begin{equation}\label{3.4}
\rho (x_1) =\left\{
\begin{aligned}
&\frac{c}{1+(x_1-p_1)^2},&\quad-\infty&<x_1<p_1,
\\
&\hphantom{1+(x_1}0,&\quad p_1&<x_1<+\infty,
\end{aligned}
\right.
\end{equation}
the constant $c=c(p,\Phi)$ is given by
\begin{equation}\label{3.10}
c(p,\Phi)=\frac{R^2\mu^2c_3(p_2)\cos^2(|p_2 - \frac{\pi}{2}| +
R)}{8\big(2\mu^2
R^2c_3(p_2)+(8\mu^2+8+c_3(p_2))(9R^2+16\pi^2)\big)},
\end{equation}
$\mu\bydef\dist\{\Phi,\mathbb{Z}\}$, $c_2(p_2)$ is the same as
in (\ref{3.13}).
\end{lemma}

The proof of this lemma is the same as the one of
Lemma~\ref{hardy1}. It is also based on similar calculations of
\cite[Sec. 7.1]{EK}, where the inequality (\ref{3.11}) was
proven for Aharonov-Bohm field. Here one also needs to replace
the function $\phi$ in \cite[Eq. (3.28)]{EK} by the function
$\varphi$ defined in (\ref{3.12}) with $p_1^-=p_1$.

\subsection{A one-dimensional model}

In this section we will show that the inequality (\ref{4.1})
holds true if the one-dimensional Schr\"odinger operator
$-\frac{d^2}{dx_1^2} + V$ in $\Ltwo (\real)$ with certain
potential $V$ is non-negative.  We will consider the case of a
compactly supported field and the Aharonov-Bohm field
simultaneously.

In view of Lemmas~\ref{hardylemma} and \ref{hardylemma2} we have
\begin{align*}
\|(-\imunit\nabla+A)u\|^2-\|u\|^2=&
\frac{1}{2}\left(\|(-\imunit\nabla+A)u\|^2-( g\, u,u)\right)
\\
 & + \frac{1}{2}\,
\|(-\imunit\nabla+A)u\|^2 +\frac{1}{2}\, (( g-2)\, u,u)
\nonumber
\\
\geq& \frac 12\, \|(-\imunit\nabla+A)u\|^2 +\frac{1}{2}\,
((\rho+ g-2)\, u,u) \, ,\nonumber
\end{align*}
where $g$ is given by (\ref{g}). Here $\rho$ is determined by
(\ref{3.2}) in the case of a compactly supported field and by
(\ref{3.4}) in the case of the Aharonov-Bohm field. Thus,
\begin{equation}\label{4.2}
\begin{aligned}
\inf_{
\genfrac{}{}{0 pt}{1}{\|u\| = 1}{u \in \domain (M_0)}
}&\left(\|(-\imunit\nabla+A)u\|^2-\|u\|^2\right)
\\
&\ge \frac{1}{2}\inf_{
\genfrac{}{}{0 pt}{1}{\|u\| = 1}{u \in \domain (M_0)}
}\left(\|(-\imunit\nabla+A)u\|^2+((\rho+g-2)\,u,u)\right).
\end{aligned}
\end{equation}
By the diamagnetic inequality (\ref{diam}) we have
\begin{equation*}
\begin{aligned}
\inf_{
\genfrac{}{}{0 pt}{1}{\|u\| = 1}{u \in \domain (M_0)}
}&\left(\|(-\imunit\nabla+A)u\|^2-\|u\|^2\right)
\\
\ge&\ \frac{1}{2}\inf_{
\genfrac{}{}{0 pt}{1}{\|u\| = 1}{u \in \domain (M_0)}
} \left(\|\nabla|u|\|^2+((\rho+g-2)\,u,u)\right)
\\
=& \ \ \frac{1}{2}\inf_{
\genfrac{}{}{0 pt}{1}{\|u\| = 1}{u \in \domain (M_0)}
} \left(\|\nabla u\|^2+((\rho+g-2)\,u,u)\right)
\\
=& \ \ \frac{1}{2}\inf_{
\genfrac{}{}{0 pt}{1}{\|u\| = 1}{u \in \domain (M_0)}
} \Bigg(\int_\Om\left(|\p_{x_1}u|^2+|\p_{x_2}u|^2\right)\D x
\\
&\hphantom{\frac{1}{2}\inf_{
\genfrac{}{}{0 pt}{1}{\|u\| = 1}{u \in \domain (M_0)}
}\Bigg(}+((\rho+g-2)\,u,u)\Bigg).
\end{aligned}
\end{equation*}
Using now (\ref{trin}) we arrive at
\begin{equation*}
\begin{aligned}
\inf_{
\genfrac{}{}{0 pt}{1}{\|u\| = 1}{u \in \domain (M_0)}
}&\left(\|(-\imunit\nabla+A)u\|^2-\|u\|^2\right)\ge
\\
&\ge \frac{1}{2}\inf_{
\genfrac{}{}{0 pt}{1}{\|u\| = 1}{u \in \domain (M_0)}
} \left(\|\p_{x_1} u\|^2+(\rho\,u,u)+2((g-1)\,u,u)\right) .
\end{aligned}
\end{equation*}
In order to establish the inequality (\ref{4.1}) it is therefore
enough to show that
\begin{equation*}
\int_0^{\pi}\, \left[\int_{\real}
|u_{x_1}(x)|^2+\rho(x_1)|u(x)|^2+2(g(x_1)-1)| u(x)|^2 \D x_1\right
] \D x_2 \geq 0,
\end{equation*}
which is equivalent to the inequality
\begin{equation} \label{onedimform}
\int_{\real}\left( |v'|^2+\rho|v|^2+2(g-1)|v|^2\right)
   \D x_1 \geq 0,
\end{equation}
for all $v\in C_0^{\infty}(\real)$. In other words, to prove
Theorems~\ref{th1}~and~\ref{th3} it is sufficient to show that
the one-dimensional Schr\"odinger operator
\begin{equation*}
-\frac{\D^2}{\D x_1^2}\, +\rho +2(g-1)
\end{equation*}
is non-negative in $\Ltwo(\real)$. The proof of this fact is the
main subject of the next section.

\subsection{The proofs of Theorems~\ref{th1}~and~\ref{th3}}

As it has been shown in the previous section to prove the
absence of the eigenvalues it is sufficient to check the
inequality (\ref{onedimform}). Due to the definition of $g$ it
can be rewritten as
\begin{equation}\label{4.3}
\int_\real |v'(t)|^2+\rho(t)|v(t)|^2\D t\ge
\frac{3}{2}\int_{-l}^l |v(t)|^2\D t.
\end{equation}
Let us show that under the assumptions of Theorems~\ref{th1},
respectively~\ref{th3} this inequality holds true. We will show
it in detail for the case of compactly supported field only (\ie
for Theorem~\ref{th1}); the case of the Aharonov-Bohm field is
similar.

We introduce a function
\begin{equation}\label{4.5}
\phi_-(t)\bydef \left\{
\begin{aligned}
&c_-\left(\frac{\pi}{2}+\arctan(t-p_1^-)\right),&&t<p_1^-,
\\
&\frac{\pi c_-}{2},&&t\ge p_1^-.
\end{aligned}\right.
\end{equation}
We remind that $c_-$ and $p_1^-$ are given by (\ref{3.2}).
Clearly, $\phi'_-(t)=\rho(t)$ for $t<p_1^-$ and $\phi'_-(t)=0$
if $t\ge p_1^-$. Keeping these properties in mind for each
$t\in(-l,l)$ we deduce the obvious equality
\begin{align*}
\frac{\pi c_-}{2} v(t)=\phi_-(t)
v(t)&=\int_{-\infty}^t\left(\phi_-(s)v(s)\right)'\D s
\\
&= \int_{-\infty}^{p_1^-}\rho(s)v(s)\D
s+\int_{-\infty}^{t}\phi_-(s)v'(s)\D s,
\end{align*}
where we also employ the fact that by the assumption of
Theorem~\ref{th1} we have $p_1^-<-l$. The equality obtained,
definition of $\phi_-$ and
Cauchy-Schwarz 
inequality give rise to an estimate
\begin{equation}\label{4.6}
\begin{aligned}
&\frac{\pi^2 c_-^2}{4} |v(t)|^2\le
2\left(\left|\int_{-\infty}^{p_1^-}\rho(s)v(s)\D
s\right|^2+\left|\int_{-\infty}^{t}\phi_-(s)v'(s)\D
s\right|^2\right)
\\
&\le 2\left(\int_{-\infty}^{p_1^-}\rho(s)\D
s\int_{-\infty}^{p_1^-}\rho(s)|v(s)|^2\D
s+\int_{-\infty}^{t}\phi^2_-(s)\D s\int_{-\infty}^{t}|v'(s)|^2\D
s\right)
\\
&\le 2\left(\frac{\pi
c_-}{2}\int_{-\infty}^{p_1^-}\rho(s)|v(s)|^2\D
s+\int_{-\infty}^{t}\phi^2_-(s)\D s\int_{-\infty}^{l}|v'(s)|^2\D
s\right).
\end{aligned}
\end{equation}
Since the function $\phi_-(t)$ is constant for $t>p_1^-$ it
follows that
\begin{align*}
\int_{-\infty}^{t}\phi^2_-(s)\D s
&=\int_{-\infty}^{p_1^-}\phi^2_-(s)\D s+\phi^2_-(p_1^-)(t-p_1^-)
\\
&=
c_-^2\int_{-\infty}^{0}\left(\frac{\pi}{2}+\arctan(s)\right)^2\D
s+\frac{\pi^2 c_-^2}{4}(t-p_1^-)
\\
&=c_-^2\pi\ln 2+\frac{\pi^2 c_-^2}{4}(t-p_1^-).
\end{align*}
Substituting the last equality into (\ref{4.6}) and using the expression
for $\phi_-(p_1^-)$ (see (\ref{4.5})) we arrive at
\begin{equation}\label{4.7}
\begin{aligned}
|v(t)|^2\le 2\Bigg(&\frac{2}{\pi
c_-}\int_{-\infty}^{p_1^-}\rho(s)|v(s)|^2\D s
\\
&+\left(\frac{4\ln
2}{\pi}+(t-p_1^-)\right)\int_{-\infty}^{l}|v'(s)|^2\D s\Bigg).
\end{aligned}
\end{equation}
In the case $c_-=0$ the fraction $\frac{1}{c_-}$ in this
inequality is understood as $+\infty$, so the inequality
valid for all possible values of $c_-$. Integration (\ref{4.7})
over $(-l,l)$ and using the obvious equality
\begin{equation*}
\int_{-\infty}^{p_1^-}\rho(s)|v(s)|^2\D
s=\int_{-\infty}^{0}\rho(s)|v(s)|^2\D s
\end{equation*}
lead us to the estimate
\begin{equation*}
\begin{aligned}
\int_{-l}^l |v(t)|^2\D t&\le 4l\Bigg(\frac{2}{\pi
c_-}\int_{-\infty}^{0}\rho(s)|v(s)|^2\D s
\\
&\hphantom{ 4l\Bigg(}+\left(\frac{4\ln
2}{\pi}-p_1^-\right)\int_{-\infty}^{l}|v'(s)|^2\D s\Bigg)
\\
&\le
\frac{4l}{\kappa_-}\left(2\int_{-\infty}^{0}\rho(s)|v(s)|^2\D
s+\int_{-\infty}^{l}|v'(s)|^2\D s\right),
\end{aligned}
\end{equation*}
where $\kappa_-$ is given by (\ref{1.10}). We can rewrite this
inequality as
\begin{equation}\label{4.8}
\kappa_-\int_{-l}^l |v(t)|^2\D t\le
4l\left(2\int_{-\infty}^{0}\rho(s)|v(s)|^2\D
s+\int_{-\infty}^{l}|v'(s)|^2\D s\right).
\end{equation}
This inequality is valid also in the case of $c_-=0$. In the
same way one can easily prove similar inequality
\begin{equation}\label{4.10}
\kappa_+\int_{-l}^l |v(t)|^2\D t\le
4l\left(2\int_{0}^{+\infty}\rho(s)|v(s)|^2\D
s+\int_{-l}^{+\infty}|v'(s)|^2\D s\right),
\end{equation}
where $\kappa_+$ is given by (\ref{1.10}).
We sum the inequalities (\ref{4.8}) and (\ref{4.10})
to get
\begin{align*}
\left(\kappa_-+\kappa_+\right)\int_{-l}^l |v(t)|^2\D t\le
4l\Bigg(&2\int_{\real}\rho(s)|v(s)|^2\D
s+\int_{-\infty}^{l}|v'(s)|^2\D s
\\
&+\int_{-l}^{+\infty}|v'(s)|^2\D s\Bigg).
\end{align*}
This implies that
\begin{equation*}
\int_{-l}^l |v(t)|^2\D t\le
\frac{8l}{\kappa}\left(\int_{\real}\rho(s)|v(s)|^2\D
s+\int_{\real}|v'(s)|^2\D s\right),
\end{equation*}
where $\kappa=\kappa_-+\kappa_+$. An immediate consequence of
the last inequality is that to satisfy (\ref{4.3}) it is
sufficient to set
\begin{equation*}
l\le \frac{\kappa}{12} ,
\end{equation*}
which coincides with the inequality (\ref{1.1}). This completes
the proof of Theorem~\ref{th1}.

The proof of Theorem~\ref{th3} is similar. One just needs to use
the inequality (\ref{4.8}) rewritten in a slightly different
way:
\begin{equation*}
\begin{aligned}
\int_{-l}^l |v(t)|^2\D t&\le 4l\Bigg(\frac{2}{\pi
c_-}\int_{-\infty}^{0}\rho(s)|v(s)|^2\D s
\\
&\hphantom{4l\Bigg(}+\left(\frac{4\ln
2}{\pi}-p_1^-\right)\int_{-\infty}^{l}|v'(s)|^2\D s\Bigg)
\\
&\le \frac{4l}{\kappa}\left(\int_{-\infty}^{0}\rho(s)|v(s)|^2\D
s+\int_{-\infty}^{l}|v'(s)|^2\D s\right),
\end{aligned}
\end{equation*}
with $\kappa$ given by (\ref{1.11}). This inequality will
immediately imply the estimate (\ref{4.3}) if the relation
(\ref{1.3}) is satisfied.

\section{Presence of eigenvalues}

In this section we will prove Theorem~\ref{th2}. We will use
the formula
\begin{equation*}
\inf\spectrum(M-1)=\inf_{
\genfrac{}{}{0 pt}{1}{\|u\| = 1}{u \in \domain (\form_M)}
}\left(\|(-\imunit\nabla+A)u\|^2-\|u\|^2\right).
\end{equation*}
If we find a test function $u\in\domain(\form_M)$ such that
\begin{equation}\label{5.2}
\|(-\imunit\nabla+A)u\|^2-\|u\|^2<0
\end{equation}
this will prove the presence of the discrete spectrum due to
Theorem~\ref{th0}. Clearly, $\domain(\form_M)$ is a subspace of
$\sobspace(\Om)$ consisting of functions that vanish on $\Gamma$.
The eigenfunction $\psi$ of $-\Delta_{\mathcal{N},\mathcal{D}}$
associated with the lowest eigenvalue $\l(l)$
belongs to $\domain(\form_M)$. We can choose this eigenfunction
being real-valued and normalized in $\Ltwo(\Om)$. Choosing
$\psi$ as a test function we have
\begin{equation}\label{5.3}
\|(-\imunit\nabla+A)\psi\|^2=\|\nabla
\psi\|^2+\|A\psi\|^2=\l(l)+\|A\psi\|^2
\le\l(l)+\max\limits_{\overline{\Om}}|A|^2.
\end{equation}
Here we used the normalization condition for $\psi$ and an
obvious relation $\l(l)=\|\nabla \psi\|^2$. By assumption the
right hand side of the last inequality is less than one, hence
the theorem is proved. Since the magnetic field $B$ determines
the magnetic vector potential $A$ up to a gauge, one naturally
should choose the potential with minimal value of
$\max\limits_{\overline{\Om}}|A|^2$; this leads us to the
inequality (\ref{1.2}).

 In conclusion let us show that the second
term on the left hand side of (\ref{1.2}) is finite. It is
sufficient to show that it is finite for some  $A$. Let $A$
be some potential associated with $B$. Since $B$ is smooth and
compactly supported, the potential $A$ can be chosen in
$C^1(\overline\Om)$. Therefore it is bounded on each bounded
subset of $\Om$. The support of $B$ is a compact set, so
there exists number $b>0$ such that $B=0$ as $x\in
\Om\setminus\Om(-b,b)$, \ie $\p_{x_2}a_2-\p_{x_1}a_1=0$ as $x\in
\Om\setminus\Om(-b,b)$. Since both domains $\Om(-\infty,-b)$ and
$\Om(b,+\infty)$ are simply connected, this immediately implies
the existence of functions $h_-\in
C^1(\overline{\Om(-\infty,-b)})$, $h_+\in
C^1(\overline{\Om(b,+\infty)})$ such that $\nabla h_-=A$ as
$x\in \overline{\Om(-\infty,-b)}$, $\nabla h_+=A$ as $x\in
\overline{\Om(b,+\infty)}$. We introduce the function
\begin{equation}\label{5.4}
h(x)=\left\{
\begin{aligned}
& h_-(x)\zeta(x_1),&x_1&<-b,
\\
&0,&-b&\le x_1\le b,
\\
& h_+(x)\zeta(x_1),&x_1&>b,
\end{aligned} \right.
\end{equation}
where $\zeta(x_1)$ is equal to one as
$|x_1|>2b$ and vanishes as $|x_1|\le b$. By definition $h\in
C^1(\overline{\Om})$. The gauge transformation $\t A\bydef A-\nabla
h$ leads us to a new vector potential $\t A$ associated
with the same field $B$. Moreover the potential $\t A$ is
compactly supported since $\nabla h=A$ if $|x_1|$ is large
enough. Since $\t A\in C^1(\overline{\Om})$, it follows that
$\max\limits_{\overline{\Om}}|\t A|^2$ is finite.

\section{Acknowledgements}

D.B. has been supported by DAAD (A/03/01031) and partially
supported by RFBR(03-01-06407) and the program ''Leading
scientific schools'' (NSh-1446.2003.1). T.E. has been supported
by ESF Programm SPECT. D.B. and T.E. thank the Stuttgart
University, where this work
has been done, for the hospitality extended to them.
Authors would like to thank T. Weidl for
permanent attention to the work and stimulating discussions.

\bigskip
\bigskip

D.~Borisov

Department of Physics and Mathematics

Bashkir State Pedagogical University

October rev. st., 3a

450000 Ufa, Russia

\bigskip
\bigskip

T.~Ekholm

Department of Mathematics

Royal Institute of Technology

Lindstedtsv\"agen 25

S-100 44 Stockholm, Sweden

\bigskip
\bigskip

H.~Kova\v r\'{\i}k

Faculty of Mathematics and Physics

Stuttgart University

Pfaffenwaldring 57

D-705 69 Stuttgart

Germany

\end{document}